\documentclass[12pt]{article}
\renewcommand{\title}{       Propagating Higgs Boundstates from Sfermions
}

\advance\textheight by \topskip
\renewcommand{\baselinestretch}{1.2}
\renewcommand{\thefootnote}{\fnsymbol{footnote}}

\newcommand{\beq}{\begin{equation}}
\newcommand{\eeq}{\end{equation}}
\newcommand{\bea}{\begin{eqnarray}}
\newcommand{\eea}{\end{eqnarray}}
\newcommand{\Label}{\label}

\begin{document}


\renewcommand{\baselinestretch}{1}
\renewcommand{\thefootnote}{\alph{footnote}}

\thispagestyle{empty}

\vspace*{-0.3cm} {\bf \hfill LBNL--42266}

\vspace*{-0.3cm} {\bf \hfill September 98} \vspace*{1.5cm}
{\Large\bf \begin{center} \title \end{center}}
{\begin{center}

\vspace*{0.5cm}
   {\begin{center} {\large\sc
                Richard Dawid\footnote{\makebox[1.cm]{Email:}
                                  Richard.Dawid@thsrv.lbl.gov}}

    \end{center} } 
\vspace*{0cm} {\it \begin{center}
    Theoretical Physics Group \\
    Ernest Orlando Lawrence Berkeley National Laboratory \\
    University of California, Berkeley, California 94720, USA
    \end{center} }
\vspace*{1cm}
\end{center}

{\Large \bf \begin{center} Abstract \end{center} }

A model of supersymmetric dynamical electroweak breaking with propagating
sfermionic Higgs boundstates is constructed. The low energy effective theory
is represented by a slight extension of the MSSM, including 2 additional
Higgs doublets and neutrino Yukawa couplings. A large $tan\beta$ is a
necessary condition. The model could be relevant in approaches which derive
propagating Higgs boundstates from strings.

\renewcommand{\baselinestretch}{1.2}

\newpage
\renewcommand{\thefootnote}{\arabic{footnote}}
\setcounter{footnote}{0}



In the framework of the minimal supersymmetric standard model (MSSM) 
the renormalization group running of the Yukawa couplings 
leads to Landau poles around the Planck scale for small or very large 
$tan\beta$. One possible way to interpret this fact is to understand the 
Higgs fields as propagating boundstates \cite{bhl, cv, SBHL} 
that break up at the Planck scale.
However, the separation of binding scale and electroweak scale is a
major problem in a supersymmetric model of dynamical electroweak breaking. 
Due to the cancellation of quadratic contributions
in SUSY it is not possible to produce a critical effective coupling from an
underlying gauge interaction. It is possible to get the 
appropriate operators from supergravity \cite{dr}, however those operators 
are highly nonlinear and do not look too attractive. The most promising
framework to produce the wanted enhanced operators seems to be 
a strongly coupled stringy scenario with some large compactified dimensions
\cite{d}. There exist two alternatives to construct dynamical
electroweak breaking in a supersymmetric framework. The possibility that
has been considered so far \cite{SBHL} is dynamical breaking induced by a
enhanced nonrenormalizable D--term interaction. This is a 
top--condensation scenario where the two effective Higgs doublets of the
MSSM--like low energy theory are 
represented by a two--sfermion boundstate and a more complicated 
boundstate dominated by a top(bottom)--antitop contribution.
General arguments in a stringy scenario however favour the second alternative,
enhanced F--terms \cite{d}. Therefore the goal of this letter 
is to formulate a model of F--term induced Higgs boundstates along the lines
followed in  the D--term induced case.

The basic idea of an effective model of dynamical electroweak breaking is to
avoid fundamental Higgs fields and instead introduce a nonrenormalizable 
coupling term. If this coupling term is stronger than a critical coupling 
value, it induces 
electroweak breaking, effectively described by a Nambu--Jona-Lasinio type
gap equation \cite{nambu}. A finetuned model of electroweak breaking 
separates the binding 
scale of the new interaction term, represented by a high energy cutoff 
$\Lambda$ in the 
effective theory, from the electroweak scale. The Nambu--Jona-Lasinio
description is an excellent approximation in this case.  In a supersymmetric 
model the separation of scales is realized by a scale difference between
$\Lambda$ and the soft breaking scale $\Delta$ plus
an enhancement of the nonrenormalizable coupling G from its natural value 
$1/\Lambda^2$ to $1/\Delta^2$ \cite{SBHL}.

An important role in finetuned dynamical electroweak breaking is played by the
renormalization group approach \cite{bhl}. This approach uses the fact that a model
of finetuned dynamical electroweak breaking is well described up to the
cutoff scale by a low 
energy effective theory with an electroweak breaking scalar sector 
(e.g. the SM, MSSM, etc.). Therefore it is possible to define the low energy 
characteristics of dynamical breaking by identifying the low energy Lagrangian
with the fundamental Lagrangian of the model at the cutoff scale. This is most
easily done by writing the nonrenormalizable interaction of the
fundamental Lagrangian in an auxiliary field formalism. The auxiliary 
fields -- modulo some normalization factor -- can then be directly identified
with the scalars of the effective theory. This 
identification gives certain conditions for the renormalization group running
of the parameters of the low energy effective theory (the so called
constituent conditions). As it is argued in detail
in \cite{bhl, bdm}, provided there is 
no new physics below the cutoff scale an electroweak breaking low energy 
theory that fulfills these conditions necessarily implies dynamical 
symmetry breaking induced by new physics at the cutoff scale. 
Our discussion will take place entirely in the framework of the renormalization
group approach. We will write down the auxiliary field Lagrangian of F--term
induced dynamical electroweak breaking, show, that the constituent conditions 
are fulfilled and extract the basic properties of the model.

For reasons of comparison we start with D--term induced electroweak breaking 
(SUSY top--condensation). There the Lagrangian has the form:

\bea
{\cal L}_D={\cal L}_{YM}&+ &\int d^{2}\theta d^{2}\bar{\theta}
(\bar{Q}e^{2V_{Q}}Q+T^{c}e^{-2V_{T}}\bar{T}^{c}+B^{c}e^{-2V_{B}}\bar{B}^{c})
(1-\Delta^{2}\theta^{2}\bar{\theta}^{2})\nonumber\\
&+& G_T\int d^{2}\theta d^{2}\bar{\theta}[(\bar{Q}~\bar{T}^{c})e^{2V_{Q}-2V_T}
(QT^{c})](1-2\Delta^{2}\theta^{2}\bar{\theta}^{2}+
\delta\bar{\theta}^{2}+\delta\theta^{2})~,
\label{Lsbhl}
\eea

where ${\cal L}_{YM}$ contains the usual SUSY kinetic terms for gauge fields,
$Q$ ($T^{c},B^c$) are SU(2) doublet (singlet) chiral quark
superfields. $\Delta^2$ and $\delta$ are SUSY soft breaking
parameters. Throughout this work superfields will be denoted by 
capital letters and 
component fields by small letters except for the vectorfield
which is identifiable by its Lorentz index. 
Reformulated in auxiliary fields eq.(\ref{Lsbhl}) corresponds to:

\bea
{\cal L}_D={\cal L}_{YM}&+&\int d^{2}\theta d^{2}\bar{\theta}
(\bar{Q}e^{2V_{Q}}Q+T^{c}e^{-2V_{T}}\bar{T}^{c}+B^{c}e^{-2V_{B}}\bar{B}^{c})
(1-\Delta^{2}\theta^{2}\bar{\theta}^{2})\nonumber\\
&+&\int d^{2}\theta d^{2}\bar{\theta}~\bar{\hat{H}}_{1}
e^{2V_{H_{1}}}\hat{H}_{1}
(1-M_{H}^{2}\theta^{2}\bar{\theta}^{2})\nonumber\\
&-&\int d^2\theta\epsilon_{ij}(\mu_{0}
\hat{H}_{1}^{i}\hat{H}_{2}^{j}(1+\hat{B}\theta^{2})-
\hat{g}_T\hat{H}_{2}^{j}Q^{i}T^{c})\nonumber\\
&-&\int d^2\bar{\theta}\epsilon_{ij}
(\mu_{0}\bar{\hat{H}}_{1}^{i}\bar{\hat{H}}_{2}^{j}
(1+\hat{B}\bar{\theta}^{2})-\hat{g}_T
\bar{T}^{c}\bar{Q}^{i}\bar{\hat{H}}^{j}_{2})  ~,
\Label{Lauxsbhl}
\eea 

with $M_{H}^{2}=2\Delta^{2}+\delta^{2}$, $\hat{B}=-\delta$, 
$V_{H_1}=V_{Q}-V_T$ and $G_T=\frac{\hat{g}_T^{2}}{\hat{\mu}^{2}}$. 
The hat will 
always denote auxiliary fields respectively their couplings. 
As already mentioned the auxiliary fields in
eq.(\ref{Lauxsbhl}) represent the two MSSM Higgs superfields at the 
cutoff scale $\Lambda$ in a different normalization.
Note that not all kinetic terms of the
composite scalar fields vanish at the cutoff scale. The auxiliary field
$\hat{H}_1$ has a kinetic term. This kinetic term is unavoidable in an
auxiliary field formulation of a supersymmetric four--fermion
theory for the following reason: Describing a supersymmetric 
four--fermion interaction requires supersymmetric auxiliary field
terms. These have to include fermion
couplings to the auxiliary scalar as well as an auxiliary scalar mass term.
Scalar mass terms are gained from the $\mu$--term
$\int d^2\theta\epsilon_{ij}\mu_{0}\hat{H}_{1}^{i}\hat{H}_{2}^{j}$
by integrating out $\hat{F}_H$--fields. For integrating out 
one needs $\hat{F}_H^2$--terms which
come from the same superfield as the scalar kinetic term. Thus a scalar
kinetic term is necessary for implementing the auxiliary field formalism for
a four fermion interaction.
Of course the second auxiliary kinetic term must not occur in order to keep the
auxiliary character of the H--fields. In the dynamical picture this relates
to the fact that the second
scalar is not a fermion-boundstate but a two--scalar boundstate and is 
produced directly by using the
auxiliary field $\hat{F}_{H_2}$ as a Lagrangian multiplier. 

To get a better understanding of the structure of the auxiliary field concept,
it is instructive to have a more general look at the possibilities
to construct an auxiliary field Lagrangian: We start with an MSSM--like
Lagrangian (forgetting about gauge fields and 
soft breaking terms at the moment)
which involves two Higgses, both possessing kinetic terms and Yukawa couplings.

\bea
{\cal L}&=&\int d^{2}\theta d^{2}\bar{\theta}~(\bar{{H}}_1H_{1}+
\bar{H}_2H_{2})
-\int d^2\theta(\epsilon_{ij}(\mu 
H_{1}^{i}H_{2}^{j}-
g_TH_{2}^{j}Q^{i}T^{c})-
g_{B}H_{1}^{j}Q^{i}B^{c})\nonumber\\
&-&\int d^2\bar{\theta}(\epsilon_{ij}
(\mu\bar{H}_{1}^{i}\bar{H}_{2}^{j}
-g_T\bar{T}^{c}\bar{Q}^{i}\bar{H}^{j}_{2})
-g_B\bar{B}^{c}\bar{Q}^{i}\bar{H}^{j}_{1})  ~.
\label{mslh}
\eea

Now we want to integrate out the massive fields $H_1$ and $H_2$.
To do this in superfield formalism we write the D--terms in
eq.(\ref{mslh}) as F--terms

\bea
\int d^{2}\theta d^{2}\bar{\theta}~\bar{H}_{1}H_{1}&=&
\int d^{2}\theta \bar{D}_{\alpha}^2\bar{H}_{1}H_{1} \nonumber\\
\int d^{2}\theta d^{2}\bar{\theta}~\bar{H}_{2}H_{2}&=&
\int d^{2}\theta \bar{D}_{\alpha}^2\bar{H}_{2}H_{2}
\label{df}
\eea

and are able to integrate out the $H$--fields under the integral 
$\int d^{2}\theta$. 
The Euler--Lagrange equations are

\bea
H_1&=&\frac{g_T}{\mu}
(QT^c-\frac{1}{4}\bar{D}_{\alpha}^2\bar{H}_2)
\Label{elhg1}\\
H_2&=&\frac{{g}_B}{{\mu}}
(QB^c-\frac{1}{4}\bar{D}_{\alpha}^2\bar{{H}}_1)
\Label{elhg2}
\eea

Re-insertion of ${H}_1$ and ${H}_2$ 
eliminates these fields up to order 
$\frac{1}{\mu^2}$. The resulting effective Lagrangian has the form

\bea
{\cal L}_{eff} &=&\int d^{2}\theta d^{2}
\bar{\theta}[G_T(\bar{Q}\bar{T}^{c})
(QT^{c})+G_B(\bar{Q}\bar{B}^{c})(QB^{c})] \nonumber\\
&+&G\int d^{2}\theta QB^cQT^c+
G\int d^{2}\bar{\theta}~
\bar{Q}\bar{B}^c\bar{Q}\bar{T}^c+ 
\mbox{higher orders}~,
\Label{esl}
\eea

where we have rewritten the SUSY derivatives $D_{\alpha}$ as 
$\theta$--integrals again and used 
$G= \frac{g_Tg_B}{\mu}$, $G_T= 
\frac{g_T^2}{\mu}$ and $G_B= 
\frac{g_B^2}{\mu}$. We get F--terms stemming from the $\mu$--term and
D--terms stemming from the kinetic terms of $\hat{H}_{1(2)}$ 
The propagation of the heavy $H$--fields is secluded in the
SUSY derivatives of the higher order contributions.
Now we remove one kinetic term in eq.(\ref{mslh}) and get a Lagrangian

\bea
{\cal L}^{\prime}&=&\int 
d^{2}\theta d^{2}\bar{\theta}~\bar{\hat{H}}_{1}\hat{H}_{1}
-\int d^2\theta\epsilon_{ij}(\hat{\mu}\hat{H}_{1}^{i}\hat{H}_{2}^{j}-
\hat{g}_T\hat{H}_{2}^{j}Q^{i}T^{c})-
\hat{g}_B\hat{H}_{1}^{j}Q^{i}B^{c})\nonumber\\
&-&\int d^2\bar{\theta}\epsilon_{ij}
(\hat{\mu}\bar{\hat{H}}_{1}^{i}\bar{\hat{H}}_{2}^{j}
-\hat{g}_T\bar{T}^{c}\bar{Q}^{i}\bar{\hat{H}}^{j}_{2})
-\hat{g}_B\bar{B}^{c}\bar{Q}^{i}\bar{\hat{H}}^{j}_{1})  ~.
\label{eshl}
\eea

Eq.~(\ref{elhg1}) reduces to

\bea
\hat{H}_1&=&\frac{\hat{g}_T}{\hat{\mu}}QT^c~.
\eea

Therefore there cannot exist any effective operators of dimension higher
than 6, the contributions of propagation for {\em both} Higgs superfields
vanish. We are confronted with an auxiliary field Lagrangian 
(which is why the $H$ fields suddenly got hats), the kinetic
term of $\hat{H}_1$ is just responsible for the derivative couplings in the
D--term. The resulting interaction Lagrangian is

\bea
{\cal L}^{\prime}=G_T\int d^{2}\theta d^{2}\bar{\theta}[(\bar{Q}\bar{T}^{c})
(QT^{c})]+G\int d^{2}\theta QB^cQT^c+G\int d^{2}\bar{\theta}
\bar{Q}\bar{B}^c\bar{Q}\bar{T}^c~.
\label{esl+-}
\eea

Now we can set to zero the second
kinetic term in eq.~(\ref{eshl}) which leaves us with a purely holomorphic
Lagrangian

\bea
{\cal L}^{\prime\prime}=
&-&\int d^2\theta\epsilon_{ij}(\hat{\mu}\hat{H}_{1}^{i}\hat{H}_{2}^{j}-
\hat{g}_T\hat{H}_{2}^{j}Q^{i}T^{c})-
\hat{g}_B\hat{H}_{1}^{j}Q^{i}B^{c})\nonumber\\
&-&\int d^2\bar{\theta}\epsilon_{ij}
(\hat{\mu}\bar{\hat{H}}_{1}^{i}\bar{\hat{H}}_{2}^{j}
-\hat{g}_T\bar{T}^{c}\bar{Q}^{i}\bar{\hat{H}}^{j}_{2})
-\hat{g}_B\bar{B}^{c}\bar{Q}^{i}\bar{\hat{H}}^{j}_{1})
\label{eshl2}
\eea

that of course corresponds to a holorphic interaction Lagrangian, i.e. a
Lagrangian without any $D$--term: 

\bea
{\cal L}^{\prime\prime}=
G\int d^{2}\theta QB^cQT^c+G\int d^{2}\bar{\theta}
\bar{Q}\bar{B}^c\bar{Q}\bar{T}^c
\label{esl2}
\eea

The other possibility is to forbid one Yukawa term which 
corresponds to forbidding the four-superfield F--terms and brings us back to
the SUSY top condensation Lagrangian of the type eq.~(\ref{Lauxsbhl}).
If we forbid both, kinetic terms for $H$s and one Yukawa term we get zero
for obvious reasons.

The Lagrangian of eq.(\ref{esl2}) plus appropriate soft breaking terms 
is exactly the interaction Lagrangian we 
are looking for. The next step is to check whether this Lagrangian can be 
identified with the MSSM Lagrangian at some scale $\Lambda$ by applying 
the constituent conditions. One can translate the vanishing of kinetic terms
in the auxiliary Lagrangian into the low energy effective theory (the MSSM
modulo possible extensions) by defining $H=g_Y\hat{H}$ and demanding a pole
for the Yukawa coupling $g_Y$ in the MSSM Lagrangian. 

At this point we have to look more carefully at the role of the Landau pole
in this framework:
Of course, if we admit the small Yukawa couplings in our effective picture, 
eventually the top Yukawa pole will drive all Yukawa couplings into the 
same pole. However this effect is due to loop contributions which involve the
Higgs doublet. Since the Higgs is a boundstate, in the full theory these 
contributions are a secondary effect compared to pure fermionic loop 
contributions. If one wants to have any control over the strong dynamics
of electroweak breaking, one has to require that these secondary effects
are suppressed in the strong coupling regime of the full theory and therefore 
irrelevant for the structure of dynamical breaking. This requirement is 
equivalent to demanding the validity of a $1/N_{c^{\prime}}$ expansion.
(See the comments on $1/N_{c^{\prime}}$ later on.) 
In the large $N_{c^{\prime}}$ limit the Higgs--loop contributions vanish.
We can therefore distinguish between poles which still exist in the
large $N_{c^{\prime}}$ limit and poles which do not. Only the first type 
constitutes a constituent condition in a model of dynamical symmetry breaking.
And we will always refer to the first type when mentioning a Landau pole 
henceforth.

In the case of SUSY top--condensation,
where only one kinetic term vanished, there was just one 
Landau pole for ${g}_T$. Now, if we want to identify the low energy 
effective Lagrangian with the Lagrangian in eq.(\ref{eshl2}) at the scale
$\Lambda$, we need two Landau poles 
for the Yukawa couplings ${g}_B$ and ${g}_T$ respectively. This however 
requires similar values for the two Yukawa couplings at the electroweak scale,
which again implies a large $tan\beta$ to produce the correct mass ratio 
$m_t/m_b$. Large values for $tan\beta$ lead to a Landau pole for the
Yukawa couplings around the Planck scale since the contributions of 
a large ${g}_B$ additionally increase the coupling values \cite{p}. Thus a
large $tan\beta$ scenario is compatible with a constituent Higgs picture.
A more detailed picture of the running behaviour of the Yukawas depends on 
the SUSY breaking structure and the embedding into a GUT which to consider is 
beyond the scope of this letter.

An additional complication arises concerning the leptonic sector.
In top condensation models usually the top is the only SM
fermion that gets its mass directly from the dynamical breaking procedure,
all other fermion masses, for quarks as well as for leptons, 
are fed down from the 
top mass in some not further specified pattern of higher order corrections. 
This seems to be plausible to
some extent since the other fermion masses are altogether considerably 
suppressed against the top mass. In our new scenario however, the bottom
quark gets its mass directly from dynamical breaking as well, the mass
difference between top and bottom is now caused by a specific form of the
vacuum and not by a different status in the process of symmetry 
breaking. This is in principle the more natural approach since it does not 
depend on treating differently two fields of the same generation. 
However it becomes
highly implausible now to treat the $\tau$--mass as a higher order effect
compared to the bottom mass since both not only belong to the same generation
but also have approximately the same value. Therefore it seems to be most 
natural to introduce an additional leptonic nonrenormalizable F--term to
get the $\tau$--mass directly from dynamical breaking as well. To do so, 
we have to couple the $\tau$--superfield to some partner to get an 
F--term that preserves hypercharge. One could couple it
to the top supermultiplet again but the much more elegant choice is to couple
it to the $\tau$--neutrino avoiding unnecessary asymmetries in the structure 
of the theory. The missing of an 
effective neutrino Yukawa coupling seems even more unnatural in the light of
the recent experimental neutrino mass signatures.
If the neutrinos are massive, there should exist
low energy neutrino Yukawa couplings which produce Dirac masses at the  
electroweak scale plus high scale right handed Majorana masses leading
to suppressed neutrino mass eigenstates via the a see--saw mechanism.
The $\tau$--neutrino Yukawa coupling naturally should have a value similar to 
the other third generation Yukawa couplings and therefore in our scenario
produce another independent Landau pole respectively a fourth Higgs boundstate.
In the large $tan\beta$ region a large neutrino Yukawa coupling does not
have a big influence on the Yukawa unification properties \cite{rs}.
Thus we define the interaction structure of our theory by the auxiliary 
Lagrangian

\bea
{\cal L}_F&=&{\cal L}_{YM}+\int d^{2}\theta d^{2}\bar{\theta}
(\bar{Q}e^{2V_{Q}}Q+T^{c}e^{-2V_{T}}\bar{T}^{c}+B^{c}e^{-2V_{B}}\bar{B}^{c})
(1-\Delta^{2}\theta^{2}\bar{\theta}^{2})\nonumber\\
&-&\int d^2\theta\epsilon_{ij}(\hat{\mu}\hat{H}_{1}^{i}\hat{H}_{2}^{j}(1+\hat{B}\theta^{2})-
\hat{g}_T\hat{H}_{2}^{j}Q^{i}T^{c}-
\hat{g}_B\hat{H}_{1}^{j}Q^{i}B^{c})\nonumber\\
&-&\int d^2\bar{\theta}\epsilon_{ij}(\hat{\mu}\bar{\hat{H}}_{1}^{i}\bar{\hat{H}}_{2}^{j}
(1+\hat{B}\bar{\theta}^{2})-\hat{g}_T\bar{T}^{c}\bar{Q}^{i}\bar{\hat{H}}^{j}_{2}-\hat{g}_B\bar{B}^{c}\bar{Q}^{i}\bar{\hat{H}}^{j}_{2})\nonumber\\
&-&\int d^2\theta\epsilon_{ij}
(\hat{\mu}^{\prime}\hat{H}_{3}^{i}\hat{H}_{4}^{j}(1+\hat{B}^{\prime}
\theta^{2})-
\hat{g}_N\hat{H}_{4}^{j}L^{i}N^{c}-
\hat{g}_E\hat{H}_{3}^{j}L^{i}E^{c})\nonumber\\
&-&\int d^2\bar{\theta}\epsilon_{ij}
(\hat{\mu}^{\prime}\bar{\hat{H}}_{3}^{i}\bar{\hat{H}}_{4}^{j}
(1+\hat{B}^{\prime}\bar{\theta}^{2})-
\hat{g}_N\bar{N}^{c}\bar{L}^{i}\bar{\hat{H}}^{j}_{4}-
\hat{g}_E\bar{E}^{c}\bar{L}^{i}\bar{\hat{H}}^{j}_{3})\nonumber\\
\label{alf}
\eea

The most general interaction Lagrangian would include also Yukawa terms of the
type $\hat{g}^{\prime}_E\hat{H}_{1}^{j}L^{i}E^{c}, 
\hat{g}^{\prime}_B\hat{H}_{3}^{j}Q^{i}B^{c}, 
\hat{g}^{\prime}_T\hat{H}_{4}^{j}Q^{i}T^{c},
\hat{g}^{\prime}_N\hat{H}_{2}^{j}L^{i}N^{c}$ (or, equivalently, the
$\mu$--terms $\hat{\mu}^{\prime\prime}\bar{\hat{H}}_{1}^{i}
\bar{\hat{H}}_{4}^{j}$ and
$\hat{\mu}^{\prime\prime\prime}\bar{\hat{H}}_{2}^{i}\bar{\hat{H}}_{3}^{j}$). 
These terms would correspond to mixed lepton--quark four--superfield couplings.
All arguments of the following discussion apply also in this more 
general case.
We will stick to the simplest case for the sake of transparency. 
The introduction of trilinear
soft breaking terms in eq.(\ref{alf}) would just correspond to a different 
parameterization of 
the same four--superfield interaction. The utility of the chosen 
parameterization will turn out later.
To identify the Lagrangian eq.(\ref{alf}) with a -- slightly extended --
MSSM Lagrangian at the cutoff $\Lambda$ we 
use four separate pole conditions for MSSM parameters. 

\bea
g_T^{-1}(\mu), ~g_B^{-1}(\mu), ~g_N^{-1}(\mu), ~g_E^{-1}(\mu) ~\rightarrow 
0 ~~~~ \mbox{for} ~~~ \mu \rightarrow \Lambda 
\label{pc}
\eea

However to really identify the two Lagrangians it is necessary that
the terms $\frac{\mu_1}{{g}_T{g}_B}{H}_1^iH_2^j$ and 
$\frac{\mu_2}{g_{E}g_{N}}H_3^iH_4^j$ do not vanish at the
cutoff scale. Otherwise the nonrenormalizable coupling G would be zero
and the dynamical picture would not be valid. In other words, the poles 
of $\mu_1$ and ${g}_T{g}_B$ respectively $\mu_2$ and $g_{E}g_{N}$
must exactly cancel. 

To check this we have to come back to a subtle
point: The full one loop renormalization group running is not an appropriate
tool to discuss the behaviour around the pole since the loop expansion 
breaks down in that region. However, as we have already mentioned, it is 
possible to discuss the pole 
region in a $1/N_{c^{\prime}}$ expansion \cite{bdl}. $N_{c^{\prime}}$
is not the number of conventional colour degrees of freedom but is 
connected to the new interaction that produces the bound states. Therefore
gluon contributions won't appear in our lowest order $1/N_{c^{\prime}}$
beta--functions. The expansion parameter $1/N_{c^{\prime}}$ of course has to
apply to the leptonic fields as well as the quark fields since they both
play the role of bound state constituents. 
$N_{c^{\prime}}$ has an obvious meaning in models
where the bound states are produced by a new gauge interaction. In models
which try to produce the bound states at the Planck scale by supergravity
or stringy effects the meaning and justification of the $1/N_{c^{\prime}}$
expansion is not that clear and eventually has to arise from the details
of such model. In any case the validity of  this expansion has to be 
assumed to have any control over the strong dynamics. 

The leading order $1/N_{c^{\prime}}$ beta
functions for our low energy model are:

\bea
\frac{d}{dt}\mu&=&\frac{1}{16\pi^2}(3g_{T}^2+3g_{B}^2)\mu \\
\frac{d}{dt}\mu^{\prime}
&=&\frac{1}{16\pi^2}(g_{N}^2+g_{E}^2)\mu^{\prime} \\
\frac{d}{dt}g_{T}&=&\frac{1}{16\pi^2}(3g_{T}^2)g_{T} \\
\frac{d}{dt}g_{B}&=&\frac{1}{16\pi^2}(3g_{B}^2)g_{B} \\
\frac{d}{dt}g_{N}&=&\frac{1}{16\pi^2}(g_{N}^2)g_{N} \\
\frac{d}{dt}g_{E}&=&\frac{1}{16\pi^2}(g_{E}^2)g_{E} 
\label{bfn}
\eea

One easily sees that $d\frac{\mu}{g_Tg_B}/dt=d\frac{\mu}{g_Ng_E}/dt=0$. 
This is the consequence of the fact that in lowest order $1/N_{c^{\prime}}$ 
the Landau poles of the Yukawa couplings as well as of the $\mu$--parameters 
are solely induced by the renormalization of the Higgs propagators. 

Next we have to take a look at the soft breaking terms. 
The low energy effective theory has soft breaking mass terms of the form

\bea
\int d^{2}\theta d^{2}\bar{\theta}
\bar{H}_ie^{2V_{H_i}}H_im^2_{H_i}\theta^2\bar{\theta}^2
\label{sbm}
\eea

with $i=1..4$. We want these terms to vanish after reparameterization at the 
scale $\Lambda$.
This translates into avoiding a pole at $\Lambda$ for the unreparameterized
term. Taking $H_1$ as an example the lowest order $1/N_{c^{\prime}}$ 
beta function is

\bea
\frac{d}{dt}m^2_{H_1}&=&
\frac{1}{8\pi^2}3g_{B}^2(m_{H_1}^2+m_B^2+m_T^2+A_1^2)
\eea

$m_T$ and $m_B$ do not run in lowest order $1/N_{c^{\prime}}$. 
The trilinear soft breaking
terms $A_i$ however are driven into a pole by the Yukawa couplings if they 
don't vanish at $\Lambda$. Thus we have the two constituent conditions

\newpage

\bea
A(\Lambda)=0 \\
-m_{H_1}^2(\Lambda)= m_B^2(\Lambda)+m_T^2(\Lambda)=2\Delta^2
\label{ccd}
\eea

where we have used the universal mass $\Delta^2$ introduced in eq.(\ref{alf}).
With these conditions fulfilled, F--term induced dynamical electroweak 
breaking in fact is a viable model.

The overall picture now shows a nonrenormalizable interaction Lagrangian

\bea
{\cal L}_F^{int}=
G\int d^{2}\theta QB^cQT^c(1+\delta(\theta^2)&+&
G\int d^{2}\bar{\theta}
\bar{Q}\bar{B}\bar{Q}\bar{T}(1+\bar{\delta}(\bar{\theta}^2) \nonumber\\ +
G^{\prime}\int d^{2}\theta LE_{\tau}L{N_{\tau}}(1+\delta^{\prime}(\theta^2)&+&
G^{\prime}\int d^{2}\bar{\theta}\bar{L}\bar{E}_{\tau}\bar{L}\bar{N}_{\tau}
(1+\bar{\delta}^{\prime}(\bar{\theta}^2)
\label{4hl}
\eea

with $\delta=-\hat{B}, ~\delta^{\prime}=-\hat{B}^{\prime}, 
~G=\frac{\hat{g}_T\hat{g}_B}{\hat{\mu}}$ and  
$G^{\prime}=\frac{\hat{g}_N\hat{g}_E}{\hat{\mu}^{\prime}}$.
The assumption of Yukawa unification would imply $G=G^{\prime}$.
The interaction structure of eq.(\ref{4hl}) 
provides dynamical electroweak breaking via a critical self 
consistence equation. The low energy effective theory is a slight extension 
of the MSSM with four Higgs doublets and neutrino Yukawa couplings. 
The parameters of the effective theory are restricted by the constituent 
conditions eq.(\ref{pc}), eq.(\ref{ccd}) and $tan\beta \approx 
m_t(\Lambda)/m_b(\Lambda)$. 
The constituent relations for the four Higgs fields are in the simplest 
case without mixing

\bea
H_1&=&\frac{\hat{g}_T}{\hat{\mu}}QT^c~ \\
H_2&=&\frac{\hat{g}_B}{\hat{\mu}}QB^c~\\
H_3&=&\frac{\hat{g}_{N}}{\hat{\mu}}LN^c~\\
H_4&=&\frac{\hat{g}_{E}}{\hat{\mu}}LE^c~
\label{cr}
\eea

In component fields this corresponds to

\bea
h_1=\frac{\hat{g}_T}{\hat{\mu}}\tilde{t}^{\dagger}\tilde{q},~etc.
\label{ccr}
\eea

As discussed earlier, in contrary to the D--term case F--term 
induced dynamical electroweak breaking cannot involve 
fermion--condensation any more, the scalar boundstates are entirely made 
of sfermions.
One can understand this model as a realization of the idea of having 
electroweak 
breaking induced by the scalar superpartners of the SM fermions. This however
does not happen via a tree level potential, but via a dynamical mechanism.

Finally one should make some remarks on prospects and experimental testability
of the described model. The phenomenology of
F--term induced dynamical electroweak breaking below the Planck scale
represents a specific choice in the parameter space of a slightly extended
MSSM. This set of parameters is similar to that suggested by some 
Yukawa unification scenarios which favour Yukawa couplings 
close to the infrared fixed point \cite{cw}. 
Thus a potential discovery of SUSY signatures could contradict or favour 
but not easily prove the discussed models. It would be interesting 
to investigate the embedding of F--term induced dynamical electroweak breaking 
into a GUT scenario, e.g. SO(10) unification. Generally GUT scenarios
that unify Yukawa couplings require a non--universal mass pattern for the 
scalar soft breaking masses at the GUT scale \cite{yuu, rs}. Since the 
conditions on the mass parameters are Planck scale conditions, this basic 
requirement 
would be fulfilled in our case. To be viable, our model needs a consistent
method to construct its enhanced non--renormalizable operators from
an underlying theory. The question whether this can be achieved most likely
has to be decided in the framework of a strongly coupled string theory.

\vspace{.5cm}
{\bf Acknowledgments:} This work was supported in part by the Director,
Office of Energy Research, Office of Basic Energy Services, of the 
U.S. Department of Energy under Contract DE-AC03-76SF00098 and in part
by the Erwin Schr\"odinger Stipendium Nr. J1520-PHY.

\newpage


\parskip=0ex plus 1ex minus 1ex


\end{document}